\documentclass[12pt,a4paper]{article} 
\usepackage{latexsym,epsf,cite}
\usepackage{amssymb,amsfonts,amsmath}
\usepackage{epsfig,epic}
\usepackage{wrapfig}
\usepackage{multirow}
\def\thefootnote{\fnsymbol{footnote}} 
\newcommand{\eq}{\begin{equation}} 
\newcommand{\en}{\end{equation}} 
\newcommand{\eqa}{\begin{eqnarray}} 
\newcommand{\ena}{\end{eqnarray}}

\newcommand{\kt}{\rangle} 
\newcommand{\lb}{\lbrack} 
\newcommand{\rb}{\rbrack} 
\newcommand{\um}{\frac12} 
\newcommand{\wt}{\widetilde} 
\newcommand{\wh}{\widehat}

\newcommand{\lan}{\langle} 
\newcommand{\ran}{\rangle}

\newcommand{\dep}{\partial}
\begin{document} 
\begin{titlepage} 
\title{Study of the 2d Ising Model with Mixed Perturbation}

\author{ 
P.~Grinza$^a$\footnote{e--mail: grinza@to.infn.it}~and 
A.~Rago$^a$\footnote{e--mail: rago@to.infn.it}\\
{\small\it $^a$ Dipartimento di Fisica  Teorica dell'Universit\`a di Torino and}\\[-0.2cm]
{\small\it Istituto Nazionale di Fisica Nucleare, Sezione di Torino} \\[-0.2cm]
{\small\it via P.Giuria 1, I-10125 Torino, Italy} }
\maketitle

\thispagestyle{empty}
 
\begin{abstract}
We study the thermodynamical observables of the 2$d$ Ising model in the neighborhood of the magnetic axis by means of numerical diagonalization of the transfer matrix. In particular, we estimate the leading order corrections to the Zamolodchikov mass spectrum and find evidence of non-vanishing contributions due to the stress-energy tensor.
\end{abstract} 
\end{titlepage} 
\setcounter{footnote}{0} 
\def\thefootnote{\arabic{footnote}} 
\section*{Introduction} 
In the past the 2$d$ Ising model has been the subject of both analytical and
numerical study.  After the original solution due to
Onsager~\cite{Onsager:1943jn} much work has been done to improve our
knowledge of the model.

It has been shown that at criticality the model can be described by a Minimal
Unitary Conformal Field Theory~\cite{bpz}.  Both magnetic and thermal
perturbations of this model have been investigated and it has been found that
these are the only two integrable perturbations of the model.  The
integrability of the model with a magnetic perturbation has been exploited by
Zamolodchikov in order to obtain the exact $S$-matrix and mass spectrum of
the theory \cite{integrable}.  Further, the exact knowledge of the $S$-matrix
has been utilized to calculate the first few terms in the spectral expansion
of the two point correlation functions via the exact calculation of the form
factors~\cite{Delfino:1995zk}. 

On the other hand, an infrared safe short distance expansion (IRS) for the
correlators has been proposed in~\cite{Guida:1995kc,Guida:1996ux}
(see~\cite{Caselle:1999mg} for a comparison of the expansions to Monte Carlo
simulations). An interesting feature of the latter approach is that, in
principle, it does not require the integrability of the model.

Apart from the analytic results there exists an extensive numerical
investigation of the model. In particular, the transfer matrix technique has
been employed in a large set of investigations: the study of the magnetic
perturbation of the model~\cite{Caselle:1999bx}, even at finite temperature
\cite{Caselle:2002cs}, the critical equation of state \cite{Caselle:2000nn}
and a classification of irrelevant operators which enters in the observables
of the theory \cite{Caselle:2001jv,Caselle:2001zd}.

However, the knowledge of the mixed perturbation of the model is still
limited.  The most important contributions to the study of this regime are
due to McCoy and Wu \cite{McCoy:ta}, Delfino et al. \cite{Delfino:1996xp} and to Zamolodchikov and Fonseca
\cite{Fonseca:2001dc}.  

This work explores this latter regime, in fact, our purpose is to investigate
the mixed perturbation model in the neighborhood of the magnetic axis.  We
apply the method developed by \cite{Caselle:1999bx} to this case.  There are
two key points of this method: the high precision numerical determination of
the eigenvalues of the transfer matrix as a function of the couplings and the
analysis of the data by means of scaling functions obtained from a CFT
approach.

The original contributions of this work can be summarized in three points.
Firstly, we verify that the CFT inspired scaling functions are in perfect
agreement with our data, moreover we are able to estimate the leading order
corrections induced by the presence of the thermal perturbation.

Secondly, we have been able to predict the existence of a term due to the
stress-energy tensor in the free energy and determine its amplitude. This is
an interesting result because the contributions of the stress-energy tensor
have been claimed to be zero if only integrable perturbations are involved.
However, this is issue is still a matter of study.

Finally we give an estimate of the corrections to the Zamolodchikov mass
spectrum and we find perfect agreement (up to some acceptable error) with
Delfino et al. \cite{Delfino:1996xp}.

This work is organized as follows.  In Section~\ref{sec:ising} we review the
standard definitions used in the Ising model and we define the normalizations
and conventions we use.  Section~\ref{sec:tm} is devoted to the explanation
of the transfer matrix technique.  The computation of the scaling functions
is illustrated in Section~\ref{quasiprimary}.  Our results are presented in
Section~\ref{sec:results}, and finally, we draw our conclusions in
Section~\ref{sec:discandconcl}.  In Appendix~\ref{numbers} we summarize all
of the known results on the amplitudes of the scaling functions, in
Appendix~\ref{scalingfunc} we present an example of the obtained scaling
functions.

\section{The Ising Model}
\label{sec:ising}
In this section we review the existing results of the 2$d$ Ising model. In
particular, we shall present the model both in its lattice form and in its
continuum formulation as a field theory.  We also define the observables that
we shall use in the following work.

\subsection{The Lattice Model}
\label{observables}
The Ising model in a magnetic field at an arbitrary temperature is defined by the partition function
\eqa
Z(\beta, h_\ell)~=~\sum_{\sigma_i=\pm 1} e^{\beta \sum_{\langle n,m \rangle}\sigma_n \sigma_m + h_\ell 
\sum_n \sigma_n },
\ena 
where the spin variable $\sigma_n$ takes the values $\pm 1$; the notation
$\langle n,m \rangle$ represents nearest neighbor sites on the lattice; the
sites are labeled by $n = (n_o,n_1)$ and the two sizes of the square lattice
are $L_0$ and $L_1$ (they are taken to be different because our transfer matrix
calculations will treat the two directions asymmetrically); the total
number of sites of the lattice will be denoted as $N = L_0 L_1$. \\
The coupling $\beta$ is the inverse of the temperature, while the magnetic
perturbation is introduced by the coupling $h_\ell\equiv H \beta$, where $H$
is the magnetic coupling. This model undergoes a second order phase
transition when $h_\ell=0$ and $\beta$ reaches its critical value $\beta_c$
\eqa \beta_c\equiv\um \log (\sqrt 2 +1) = 0.4406868 \dots \,.\ena Now we
define the observables which we shall consider in the following work.
\begin{itemize}
\item{The{\it Magnetization} per site is defined as
\eqa
M(\beta, h_\ell) = \frac1N \frac{\dep}{\dep h_\ell}(\log Z(\beta, h_\ell) )=
\frac1N \lan \sum_i \sigma_i \ran.
\ena
}
\item{The {\it Magnetic Susceptibility} is defined as 
\eqa
\chi (\beta, h_\ell) = \frac{\dep M(\beta, h_\ell)}{\dep h_\ell}.
\ena
}
\item{The {\it Free Energy} is defined as
\eqa
f(\beta, h_\ell) = \frac1N \log Z(\beta, h_\ell).
\ena
The free energy is composed of a ``bulk'' term 
$f_b(\beta, h_\ell)$, which is an analytic function of $h_\ell$ and $\beta$,
and a ``singular'' term encoding the relevant information about the 
theory in the neighborhood of the critical point. From the exact solution
of the lattice model at $h_\ell=0$ and $\beta=\beta_c$ we can compute the
value of $f_b(0,0)$
\eqa
f_b = \frac{2G}{\pi} + \um \log 2= 0.9296953982 \dots \,,
\ena 
where $G$ is the Catalan constant.
}
\item{The {\it Internal Energy}  density is defined as
\eqa
\wh{E}(\beta, h_\ell) = \frac{1}{2N} 
\lan \sum_{\langle n,m \rangle}\sigma_n \sigma_m \ran.
\ena
As for the free energy, we have both a bulk and a singular part.
If we define $E_b (\beta, h_\ell)$ as the bulk contribution, we can
obtain the value of $\epsilon_b=E_b (0,0)=\frac{1}{\sqrt 2}$
(by using Kramers-Wannier duality or CFT techniques) and we can 
also define (for future convenience)
\eqa
E (\beta, h_\ell) = \wh{E}(\beta, h_\ell) - \epsilon_b.
\ena
}
\item{{\it Time slice correlation functions.} \\
We can define the zero momentum projections of the two-point 
correlation functions $\lan \sigma(r) \sigma(0)\ran $ and
$\lan \epsilon(r) \epsilon(0)\ran $ (they are also called time slice
correlators). The magnetization of a row of the lattice (time slice)
is given by
\eqa
S_{n_0} = \frac{1}{L_1} \sum_{n_1} \sigma_{(n_0,n_1)}
\ena
hence, the correlation function between time slices is given by
\eqa
G^0_{\sigma \sigma}(n_\tau)=\sum_{n_0} \left[ \lan S_{n_0} S_{n_0 + n_\tau} \ran 
-\lan S_{n_0}  \ran^2  \right]
\ena
where label $0$ indicates that it is the zero momentum projection
of the original correlator. 
}
\end{itemize}
\subsection{The Ising Field Theory}
The 2$d$ Ising model near the phase transition can be described via a
minimal unitary Conformal Field Theory (with central charge $c=1/2$) perturbed
with the energy and magnetization densities $\epsilon(x)$ and
$\sigma(x)$
\eqa
{\mathcal A} \ = \ {\mathcal A}_{CFT} + h \int \textrm{d}^2 x~
\sigma(x) + \tau \int \textrm{d}^2 x~ \epsilon(x).
\label{action}
\ena
Where ${\mathcal A}_{CFT}$ is the action of the model at the critical temperature without external magnetic field.\\
The coupling constant $h$ represents the magnetic perturbation of the model
and, as stated before, it is the continuum version of the previously defined
coupling $h_\ell$. The other coupling constant $\tau$ represents the thermal
perturbation and, near criticality, it is proportional to the reduced
temperature $t$ \eqa t = \frac{\beta_c-\beta}{\beta_c} \ena
which we use in the rest of the paper.\\
In the following Section we give a brief report of the known results on the
field theory description of the model.
\subsubsection{The Critical Theory $(t =0,h_\ell =0)$}
The Ising model is the lowest model of the so-called ``Minimal Unitary''
series of conformal theories whose central charge is given by \eqa c_p = 1-
\frac{6}{p(p+1)}, \ \ \ \ \ \ p=\;3, \dots.  \ena 
The peculiarity of these models is that they possess a finite set of primary
fields; as a consequence the whole space of local operators of the theory can
be built by applying the generators of the Virasoro algebra to the primary
fields. The operators obtained in this way are called secondary fields or
descendants.  Following this route, one is led to organize the operator
content of the theory in conformal families, i.e. the sets of descendants of
each
primary field. \\
The operator spectrum of the Ising model consists of three primary operators
\begin{itemize}
\item{Identity $\Rightarrow$ $\Delta_I = 0$}
\item{Magnetization $\Rightarrow$ $\Delta_\sigma = 1/16$}
\item{Energy $\Rightarrow$ $\Delta_\epsilon = 1/2$}
\end{itemize}
where $X_i = 2 \Delta_i$ is the scaling dimension of each operator.

We assume for the field $\sigma (x)$ and $\epsilon (x)$ the usual CFT
normalization 
\eqa
\langle \sigma (x) \sigma (0) \rangle =  \vert x \vert^{-1/4}, \ \ \ \ \ \ \ \ \ 
\langle \epsilon (x) \epsilon (0) \rangle = \vert x \vert^{-2}
\ena
so they have scaling dimensions $1/16$ and $1/2$ respectively; the
parameters $h$ and $t$ have dimensions $15/8$ and $1$. \\
Hence there are three conformal families that descend from these operators 
and the general expression for the descendants $\mathcal{O_{[\phi]}}$ is
\eqa
{\mathcal O}_{[\phi]}= L_{-k_s\cdots} L_{-k_1} 
{\overline {L}}_{-m_p\cdots} {\overline {L}}_{-m_1}\phi 
\ena
where
\eqa
\sum_{i=1}^{s} k_i = n~;  \ \ \ \ \ \ \   \sum_{i=1}^{p} m_i 
=\overline{n},
\ena
the operator $\phi$ is one of the primary fields of the theory.\\
The scaling dimension and the conformal spin of the operators are given by
\begin{equation}
\begin{split}
X_{\mathcal{O}}& =  \Delta_{\mathcal{O}} + \overline{\Delta}_{\mathcal{O}}
=  n + \overline {n} + \Delta_\phi 
+\overline{\Delta}_\phi\\
s_{\mathcal{O}} & = n - \overline {n}. 
\end{split}
\end{equation}  
Among secondary fields, the quasi-primary fields play a special role.
A descendant field $Q$ is called quasi-primary when
\begin{itemize}
\item{$L_1 \vert {Q}  \kt = 0$}
\item{ $ \vert {Q} \kt $ is not a null vector}.
\end{itemize}
It will be shown in Section~\ref{quasiprimary} that quasi-primary fields can be used as
building blocks for the construction of an effective Hamiltonian
for the lattice model near criticality.
\subsubsection{The Ising Field Theory with Magnetic Perturbation $(t =0, h_\ell\neq 0)$}
If $\beta$ is fixed to its critical value $\beta_c=\um \log (1+\sqrt 2)$ and the magnetic field is switched on, 
the field theory is still integrable, i.e. it possesses an infinite number of
integrals of motion. This implies the exact knowledge of the $S$-matrix
and the mass spectrum. The latter result, due to Zamolodchikov \cite{Zamolodchikov:1989fp},
suggests that the model could be described by a scattering theory with
a spectrum of eight self-conjugated particles with masses
\begin{equation}
\begin{split}
\label{mass_spectrum}
m_2 &= 2 m_1 \cos\frac{\pi}{5} = (1.6180339887..) \,m_1\,,\\
m_3 &= 2 m_1 \cos\frac{\pi}{30} = (1.9890437907..) \,m_1\,,\\
m_4 &= 2 m_2 \cos\frac{7\pi}{30} = (2.4048671724..) \,m_1\,, \\
m_5 &= 2 m_2 \cos\frac{2\pi}{15} = (2.9562952015..) \,m_1\,,\\
m_6 &= 2 m_2 \cos\frac{\pi}{30} = (3.2183404585..) \,m_1\,,\\
m_7 &= 4 m_2 \cos\frac{\pi}{5}\cos\frac{7\pi}{30} = (3.8911568233..) \,m_1\,,\\
m_8 &= 4 m_2 \cos\frac{\pi}{5}\cos\frac{2\pi}{15} = (4.7833861168..) \,m_1\,
\end{split}
\end{equation}
where $m_1\equiv m_1(h)$ is the fundamental mass of the theory, and is given by
\eqa
m_1(h)={\mathcal C} h^\frac{8}{15}.
\ena
The numerical value of ${\mathcal C}$ was computed by Fateev \cite{Fateev:1993av}
\eqa
{\mathcal C}=\frac{4 \sin \frac{\pi}{5} \Gamma(\frac{1}{5})}{\Gamma(\frac{2}{3})\Gamma(\frac{8}{15})}\left(\frac{4 \pi^2 \Gamma(\frac{3}{4}) \Gamma^2 (\frac{13}{16})}{\Gamma(\frac{1}{4})\Gamma^2 (\frac{3}{16})}\right)^{\frac{4}{5}}=4.40490858\dots.
\ena
The vacuum expectation values of energy and magnetization can be
parame\-trized as 
\eqa
\langle \epsilon \rangle = A_\epsilon~h^{8/15}, \ \ \ \ \ \ \ \ \ \ \ \ 
\langle \sigma \rangle = A_\sigma~h^{1/15}\
\ena
where the amplitudes $A_\epsilon$ and $A_\sigma$ can be computed 
exactly \cite{Fateev:1997yg}
\eqa
A_\epsilon~=~2.00314 \dots, \ \ \ \ \ \ \ \ \ \ \ \
A_\sigma~=~1.27758227 \dots.
\ena
\subsubsection{Non-integrable Perturbation of Ising Model $(t \neq 0,h_\ell\neq 0)$}
The non-integrable perturbation of the 2$d$ Ising model\footnote{See, e.g. \cite{Yurov:1990kv} for a detailed discussion of the model with thermal perturbation only $(t \neq 0,h_\ell= 0)$.} was treated in \cite{Delfino:1996xp} in the framework of the Form Factors Perturbation theory.\\
The starting point of FFPT is to consider the mixed perturbation as the 
perturbation of an integrable QFT.

The action (Eq.~\ref{action}) describes a one parameter family of theories
labeled by the adimensional scaling variable $\xi$ \eqa \xi \equiv
\frac{t}{h^\frac{8}{15}}.  \ena In this framework it is possible to calculate
(at first order) the corrections to the mass spectrum and vacuum energy of
the integrable theory.  Two limits have been investigated in
\cite{Delfino:1996xp}: the magnetic perturbation of a free massive Majorana
fermion ($\xi \to \infty$) and the thermal perturbation of the integrable
magnetic perturbation of the
critical Ising model ($\xi \to 0$). \\
The detailed discussion of \cite{Delfino:1996xp} shows that in the former
case ($\xi \to \infty$), a straightforward application of the method is not an
easy task. In fact, if we are in the high temperature regime of the model,
the corrections at first order vanish.  A second order calculation is
required but is a non-trivial computation.  However, in the low temperature
regime, the
divergence remains at first order, and is the signal kink's confinement.

The latter case ($\xi \to 0$), however, is more tractable and gives rise to
some quantitative predictions of the mass spectrum of the model.  The
results are listed below
\begin{equation}
\begin{split}
\frac{\delta {\mathcal E}_{vac}}{\delta m_1 } & \simeq  -0.0558 \dots m_1^0,
 \\
\frac{\delta m_2 }{\delta m_1 } & \simeq  0.8616 \dots,
 \\
\frac{\delta m_3 }{\delta m_1 } & \simeq  1.5082 \dots.
\end{split}
\end{equation}
The non integrable perturbation gives rise to an important quantitative
difference on the masses above the threshold.  In fact, the integrability of
the pure magnetic perturbation prevents the creation of new particles at
energies above the threshold (and from any other inelastic process). The
explicit breakdown of integrability implies that particles above threshold
become unstable and are expected to decay.  This new feature manifests itself
in the corrections to the masses, namely they develop an imaginary part. This
effect has been seen explicitly in the case of the first mass above the
threshold $m_4$. The first order correction is real and given by \eqa
\frac{\delta m_4 }{\delta m_1 } & \simeq & 1.1460 \dots \ena while, at second
order, one expects  a non-zero value of $\textrm{Im} \; m_4^2 $ (it
was not computed in \cite{Delfino:1996xp} because the authors' analysis
only covers the first order contribution).
\subsection{Relation Between Lattice and Continuum Operators} 
It is useful to define, for future convenience, the relations between the 
lattice and continuum definitions of the energy and magnetization operators.
Near the critical point, the simplest choices for the lattice operators are
\begin{itemize}
\item{{\it{Spin operator}}
\eqa
\sigma_l(x)=\sigma_x
\ena
where the index $l$ indicates that it is a lattice discretization of
the continuum operator. The magnetization per site is defined as
\eqa
\sigma_l = \frac1N  \sum_x \sigma_l(x).
\ena
}
\item{{\it{Energy operator}}
\eqa
\epsilon_l = \frac14 \sigma_x \left[ \sum_{y \in \{ n.n. x \}} \sigma_y 
\right] - \epsilon_b
\ena
where the sum runs over the nearest neighbor sites of $x$ and
$ \epsilon_b$ is the bulk term. The energy per site is defined as
\eqa
\epsilon_l = \frac1N  \sum_x \epsilon_l(x).
\ena
}
\end{itemize}
Off-critical corrections to scaling of these operators
will be discussed later.
\subsection{Converting between Lattice and Continuum Units}
In order to fix the conversions between lattice and continuum units, we shall
follow the careful discussion of \cite{Caselle:1999bx}. One can 
write the lattice versions of $\sigma$ and $\epsilon$ as
\begin{equation}
\begin{split}
\sigma_l & = f_0^\sigma(t,h) \sigma + f_i (t,h) \phi_i 
 \\
\epsilon_l & = g_0^\epsilon(t,h) \epsilon + g_i (t,h) \phi_i 
\end{split}
\end{equation}
where $f_0^\sigma$, $f_i$, $g_0^\epsilon$, $g_i$ are suitable functions of 
$t$ and $h$ (which also depend on the parity properties of the operator), while
the operators $\phi_i$ are all other fields (relevant and irrelevant)
of the theory respecting the symmetries of the lattice. We also have
\eqa
h_\ell = b_0(t,h) h
\ena
which is the relation between the lattice coupling constant $h_\ell$ and
the continuum magnetic field $h$; $b_0(t,h)$ is an even function 
of $h$.

At first order in $t$ and $h$, i.e. near criticality when $t \to 0$ and
$h \to 0$, we have
\eqa
\sigma_l = R_\sigma \sigma, \ \ \ \ \ \epsilon_l = R_\epsilon \epsilon,
\ \ \ \ \ h_\ell = R_h h
\ena  
where the constants $R_\sigma$, $R_\epsilon$, $R_h$ are defined as
\begin{equation}
\begin{split}
R_\sigma & = \lim_{t,h \to 0} f_0^\sigma(t,h)  \\
R_\epsilon &= \lim_{t,h \to 0} f_0^\epsilon(t,h)  \\
R_h &= \lim_{t,h \to 0} b_0 (t,h).
\end{split}
\end{equation}
The previous normalizations were fixed in \cite{Caselle:1999bx} by comparison
with the explicit expression of the spin-spin and energy-energy
critical correlation functions on the lattice. The numerical
results are
\begin{equation}
\begin{split}
R_\sigma & = 0.83868 \dots\\
R_\epsilon &= \frac1\pi = 0.31831 \dots\\
R_h &= R_\sigma^{-1} = 1.1923 \dots.
\end{split}
\end{equation}
\section{Transfer Matrix Technique}
\label{sec:tm}
We face the problem of computing mass spectrum and observables by
numerical diagonalization of the transfer matrix. This technique,
introduced in 1941 by Kramers and Wannier \cite{kwan}, was extensively used 
by Baxter to obtain analytic solutions of certain statistical mechanical
models \cite{baxterbook}. For further details and discussions about 
numerical implementations of the
transfer matrix see, e.g. \cite{tramat}. \\
The basic idea is to rewrite the Boltzmann weight by means of the
transfer matrix $T(u_i,u_j)$
\eqa
T(u_{n_0},u_{n_0+1})  =
V(u_{n_0})^{1/2}~U(u_{n_0},u_{n_0+1})~V(u_{n_0+1})^{1/2}
\ena
with
\begin{equation}
\begin{split}
U(u_{n_0},u_{n_0+1}) & = \exp\left(\beta \; \sum_{n_1 = 1}^{L_1} \;
\sigma_{(n_0,n_1)} \sigma_{(n_0+1,n_1)} \right)\\
V(u_{n_0}) & = \exp\left(\beta \;\sum_{n_1 = 1}^{L_1} 
\sigma_{(n_0,n_1)} \sigma_{(n_0,n_1+1)} 
\;+\; h_\ell \;\sum_{n_1}^{L_1} \sigma_{(n_0,n_1)} \right) \;\;
\end{split}
\end{equation}
where $u_{n_0}=( \sigma_{(n_0,1)}, \sigma_{(n_0,2)}, \dots, \sigma_{(n_0,L_1)} )$ is
the spin configuration at the row (time-slice) $n_0$.
The previous position implies that the partition function becomes
\begin{equation}
\begin{split}
\label{prtfnt}
Z(\beta, h_\ell) & = \sum_{\sigma_i=\pm 1} e^{\beta \sum_{\langle n,m \rangle}\sigma_n \sigma_m + h_\ell 
\sum_n \sigma_n }\\
& = \sum_{\sigma_i=\pm 1}  T(u_1,u_2)\;T(u_2,u_3)  \; \cdots \; T(u_{L_0},u_1) \\
& = \textrm{tr}\;T^{L_0}~=~\sum_i \lambda_i^{L_0}
\end{split}
\end{equation}
where $T$ is a positive and symmetric $2^{L1} \times 2^{L1}$ matrix,
whose (real) eigenvalues are the $\lambda_i$.  
\subsection{Observables in Transfer Matrix Formalism}
The numerical computation of eigenvalues and eigenvectors of the
transfer matrix enables us to compute all the observables
we need, provided that we specify the values of $h_\ell$, $t$ and $L_1$.
We can derive all the observables from the partition function $Z(\beta, h_\ell)$ (\ref{prtfnt}). For the derivation see \cite{Caselle:1999bx}.
\begin{itemize}
\item{{\it{Free energy}}
\eqa
f(\beta, h_\ell) & =& \frac{1}{L_0  L_1} \log Z(\beta, h_\ell).
\ena
This expression simplifies in the limit $L_0 \to \infty$, in fact the leading
contribution is due to the maximum eigenvalue $\lambda_{max}$
\eqa
f(\beta, h_\ell) & \sim &  \frac{1}{ L_1} \log \lambda_{max}.
\ena
}
\item{{\it{Magnetization}}
\eqa
\lan \sigma_l \ran = \frac{ \mbox{tr} \; S \; T^{L_0} } { \mbox{tr} \;T^{L_0} }
\ena
where $S(u_{n_0},u_{n_1}) = \delta (u_{n_0},u_{n_1}) \sigma_{(n_0,1)} $.
In the limit $L_0 \to \infty$ we have
\eqa
\lan \sigma_l \ran = \lan 0 \vert S  \vert 0 \ran
\ena
where $ \vert 0 \ran$ is the eigenvector associated to the $\lambda_{max}$ eigenvalue.
}
\item{{\it{Energy}}
\vskip0.2cm
The case of the internal energy is analogous similar to that of magnetization.
In the limit $L_0 \to \infty$ we have
\eqa
\lan \epsilon_l \ran = \lan 0 \vert E  \vert 0 \ran
\ena
where the matrix $E(u_{n_0},u_{n_1})$ is given by
\eqa
E(u_{n_0},u_{n_1}) = \delta (u_{n_0},u_{n_1}) \sigma_{(n_0,1)} \sigma_{(n_0,2)}.
\ena
}
\item{{\it{Correlation functions and mass spectrum}}
\vskip0.2cm
In the limit $L_0 \to \infty$ the time slice correlation function
is defined as
\eqa
\lan S_0 \; S_{t}\ran  = \sum_i \exp(-m_i \; |t|) 
\lan 0|\tilde S |i \ran \; \lan i|\tilde S |0 \ran
\ena
with $\tilde S= \frac{1}{L_1} \; \delta(u_{n_0},u_{n_1}) \; \sum_{n_1} \sigma_{(n_0,n_1)} $. \\
The mass spectrum $m_i$ is given by
\eqa
m_i = - \log \left(\frac{\lambda_i}{\lambda_0}\right)
\ena
where the eigenvalues are organized in decreasing order of magnitude 
$\lambda_{max} \equiv \lambda_0 > \lambda_1 > \dots > \lambda_i > \dots $ and 
$| i \ran$ are the normalized eigenvectors of $T$.
}
\end{itemize}
\section{Scaling Functions}
\label{quasiprimary}
In order to study the collected data obtained from the TM, we need to know
the behavior of the measured operator as a function of the perturbation
variables.
The fundamental ingredient of this construction is the knowledge of the whole spectrum of operators of the theory, including the OPE between them.\\
Hence, the operator content of the 2$d$ Ising model at the critical point
previously discussed, enables us to build an effective Hamiltonian for the
perturbed model.  As discussed in detail in \cite{Caselle:1999bx}, the aim of
this effective Hamiltonian is not to describe the model at a scale comparable
with the lattice spacing; instead it has to be considered as the Hamiltonian
describing the model after a suitable number of Renormalization Group
transformations, i.e. at a scale that is larger with respect to the lattice
spacing.
\subsection{Lattice Construction of the Ising Model Via CFT Operators}
The main idea then is to use the whole spectrum of conformal operators,
defined on the continuum, to describe the corrections to scaling 
(due to the lattice) in the observables of the model.
In order to build this Hamiltonian explicitly, we have to take into account,
in principle, the following ingredients:
\begin{itemize}  
\item{{\it{Symmetries of the model}} \\
    Unlike the case of the model at the critical point, which exhibits two
    exact symmetries (${\mathbb Z}_2$ and duality), the presence of the
    magnetic field explicitly breaks all of them. Hence, in this case, there
    are no constraints coming from symmetries (it is useful to remember that
    this argument, in the critical case, selects only the fields belonging to
    the conformal family of the identity).  }
\item{{\it{Symmetries of the lattice}} \\
    It is crucial to define the geometry of lattice we are using in the
    transfer matrix calculation. In the following we consider a square
    lattice.  This means that the rotational symmetry of the CFT is broken
    down to the dihedral subgroup $D_4$ and also operators with spin are
    allowed.  Hence, the residual symmetry group (rotations of integer
    multiples of $\pi/2$) implies that only operators with spin $j=4k$, $k
    \in {\mathbb N} $, can appear on the lattice (see \cite{Caselle:1999bx}
    for a detailed discussion).  }
\item{{\it{Lattice $\leftrightarrow$ continuum relations}} \\
    Lattice operators are defined in terms of continuum operators as follows
\begin{equation}
\begin{split}
  \sigma_l & = f_0^\sigma (h_\ell,t) \sigma + h_\ell f_0^\epsilon
  (h_\ell,t)\epsilon + f_i^\sigma (h_\ell,t) \sigma_i + h_\ell f_i^\epsilon
  (h_\ell,t)\epsilon_i +
  h_\ell f_i^{\mathbf 1}  \eta_i \\
  \epsilon_l & = g_0^\epsilon (h_\ell,t)\epsilon + h_\ell g_0^\sigma
  (h_\ell,t)\sigma + h_\ell g_i^\sigma (h_\ell,t) \sigma_i + g_i^\epsilon
  (h_\ell,t)\epsilon_i + g_i^{\mathbf 1} \eta_i
\end{split}
\end{equation}
where $f$ and $g$ are functions of the reduced temperature $t$, and 
they are even functions of the magnetization $h_\ell$. Furthermore, 
the operators appearing in the previous expressions have to be 
compatible with the described symmetries.
}
\end{itemize} 
Now we are able to write the following lattice (effective) Hamiltonian
\eqa
H_{lat} = H_{CFT} + h_\ell \sigma + t \epsilon + u_i \Psi_i 
\ena
where $\Psi_i$ are the quasi-primary fields belonging to the whole set
of conformal families of the theory with spin $j=4k$, $k \in {\mathbb N}$. \\
The least irrelevant fields which enter the expression of $H_{lat}$ are
built starting from the following quasi-primary fields of the family
of the identity
\begin{equation}
\begin{split}
Q_2^{\mathbf 1} & = L_{-2} {\mathbf 1}\\
Q_4^{\mathbf 1} & = ( L_{-2}^2 - \frac{3}{5} L_{-4} ) {\mathbf 1}
\end{split}
\end{equation}
where the notation $Q^\eta_n$ is used to denote with $\eta$ the conformal
family and with $n$ the level of descent. The same can be done for all the 
other families (for a list of low-lying quasi-primary states, up to
level 10, see \cite{Caselle:2001jv}).\\
All the results are reported in Table~\ref{tab:operators}.
\begin{table}[ht]
\begin{center}
\begin{tabular}{c|c|c|c|}
\cline{2-4}&&&\\[-4mm]
&\multirow{2}{26mm}{\centerline{Spin-0 Sector}}&\multirow{2}{26mm}{\centerline{Spin-4 Sector}}&RG\\
&&&Eigenvalue\\
\cline{2-4}
\multicolumn{4}{}{}\\[-4.5mm]
\cline{2-4}
&&&\\[-4mm]
\multirow{2}{20mm}{\centerline{Identity}}&$Q_2^{\mathbf 1} {\bar Q}_2^{\mathbf 1} \equiv T \bar{T}$&$Q_4^{\mathbf 1} + {\bar Q}_4^{\mathbf 1} \equiv T^2 +  \bar{T}^2$&$-2$\\[1mm]
&$Q_4^{\mathbf 1} {\bar Q}_4^{\mathbf 1}$&$Q_6^{\mathbf 1} {\bar Q}_2^{\mathbf 1} + Q_2^{\mathbf 1} {\bar Q}_6^{\mathbf 1}$&$-6$\\
&&&\\[-4mm]
\cline{2-4}
&&&\\[-4mm]
\multirow{2}{20mm}{\centerline{Energy}}&$Q_4^{\epsilon} {\bar Q}_4^{\epsilon}$&
&$-7$\\[1mm]
&&$Q_4^{\epsilon} + {\bar Q}_4^{\epsilon}$&$-2 -\frac{1}{2}$\\
&&&\\[-4mm]
\cline{2-4}
&&&\\[-4mm]
\multirow{2}{20mm}{\centerline{Spin}}&$Q_3^{\sigma} {\bar Q}_3^{\sigma}$&&$-4-\frac{1}{8}$\\[1mm]
&$Q_5^{\sigma} {\bar Q}_5^{\sigma}$&$Q_3^{\sigma} {\bar Q}_7^{\sigma} + Q_7^{\sigma} {\bar Q}_3^{\sigma}$&$-8-\frac{1}{8}$\\[2mm]
\cline{2-4}
\end{tabular}
\end{center}
\caption{Low-lying quasi-primary operators. $T$ is the stress-energy tensor.}
\label{tab:operators}
\end{table}
\subsection{Computation of the Scaling Functions}
We are now in a position to compute the singular part of the scaling
functions of thermodynamic observables, e.g. the free energy, making use of
the lattice Hamiltonian we have constructed.  This is achieved by starting
with the partition function $Z(h_\ell, \beta)$ of the lattice Hamiltonian
$H_{lat}$.  Expanding this expression, it is possible to write down a formal
series expansion in the variables $h_\ell$ and $\xi \rightarrow 0$.  Hence we
are able to obtain the scaling functions expressions~\footnote{We report in
  appendix \ref{scalingfunc} the expression of some of the scaling
  functions.} for the non-scaling corrections to the observables of
Section~\ref{observables}.  To obtain these results, some remarks are in order:
\begin{itemize} 
\item{In the (numerical) transfer matrix analysis we are interested in the
    limit $\xi \to 0$ (the thermal perturbation is smaller than the magnetic
    one), so we can consider the following expressions for the VEV of the
    operators \eqa \lan \mathcal{O} \ran~=~ A_\mathcal{O}~ h^{\frac{dim\;
        \mathcal{O} }{dim \; h}}~q_\mathcal{O} (\xi,h_\ell) \ena where
    $q_\mathcal{O}$ is an analytic function of its arguments. In the general
    case this ansatz is not correct because the presence of resonances induce
    the appearance of logarithms in the expression of VEVs.  }
\item{In the formal expansion of the partition function $Z(h_\ell, \beta)$,
    we find that there also appear products of the conformal fields contained
    in $H_{lat}$. The correct way to deal with these is to use the fusion
    rules of the conformal theory
\begin{equation}
\begin{split}
\lb \epsilon\rb \lb\epsilon\rb & \sim \lb\mathbf{1}\rb\\
\lb\sigma\rb \lb\epsilon\rb & \sim \lb\sigma\rb \\
\lb\sigma\rb \lb\sigma\rb & \sim \lb\mathbf{1}\rb + \lb\epsilon\rb
\end{split}
\end{equation}
where the notation $\lb\dots\rb$ means that we are referring to the whole conformal family.
}
\end{itemize}
\subsubsection{A Peculiar Case: The Free Energy}
\label{4.2.1}
In this Section we show how to determine the scaling functions of the model
in the region of interest ($\xi \rightarrow 0$).  We have computed the
scaling functions by means of two different methods:
the renormalization group, and CFT approach.\\
If we consider the free energy as an example, in the renormalization group
approach, we can write it as the sum of three contributions \eqa
\label{rgeq}
f(t,h_\ell)= f_b (t,h_\ell) + f_{sing} (t,h_\ell) + f_{log} (t,h_\ell)
\ena
The bulk term takes into account analytic contributions in the
variables $t$ and $h_\ell$ due to non critical behavior
\begin{equation}
\begin{split}
\label{fbulk}
f_b (t,h_\ell) & = f^b_{0,0} + f^b_{0,2}~h_\ell^2 + f^b_{0,4}~h_\ell^4 + (f^b_{1,0} + f^b_{1,2}~h_\ell^2 + f^b_{1,4}~h_\ell^4)~t+ 
\\
& + (f^b_{2,0} + f^b_{2,2}~h_\ell^2 + f^b_{2,4}~h_\ell^4)~t^2 + (f^b_{3,0} + f^b_{3,2}~ h_\ell^2 + f^b_{3,4}~ h_\ell^4)~t^3 +O(h_\ell^6,t^4).
\end{split}
\end{equation}
Only even powers of $h_\ell$ appear because the free energy is
even under $\mathbb{Z}_2$ transformations. \\
The non-analytic contribution is given by the master equation of the RG
\eqa
\label{rgeqnonan}
f_{sing}(t,h_\ell)= g_h^{2/ \Delta} Y \left( \frac{g_t}{g_h^{1/\Delta}},
\left\{ g_u~ g_h^{ | y_u | / \Delta} \right\} \right)
\ena
where the scaling variables $g_h(t,h_\ell)$, $g_t(t,h_\ell)$, $g_u(t,h_\ell)$
are defined
in the usual way \cite{Caselle:2000nn}
\begin{equation}
\begin{split}
\label{scalvars}
g_t(t,h_\ell) &= t + b_t h_\ell^2 + c_t t^2 + d_t t^3 + e_t t h_\ell^2 +
f_t t^4 + g_t h_\ell^4 + h_t t^2 h_\ell^2 + \dots\\
g_h(t,h_\ell) &=  h_\ell (1+ c_h t + d_h t^2 + e_h h_\ell^2 + f_h t^3
+ g_h t h_\ell^2) + \dots\\
g_u(t,h_\ell) &= u + a_u t +b_u h_\ell^2 + c_u t^2 + d_u t^3 + e_u t
h_\ell^2 + f_u h_\ell^4  + l_u t^2 h_\ell^2 + \dots
\end{split}
\end{equation}
and some of the coefficients are known either exactly or numerically
(see appendix \ref{numbers}). \\
It has been known for a long time \cite{aarony} that, in order to take into
account the logarithmic divergence of the specific heat of the Onsager
solution \cite{Onsager:1943jn}, the term $f_{log}$ (\ref{rgeq}) must have the
following form \eqa f_{log}=g_t^2(t,h_\ell) \log
\left(|g_t(t,h_\ell)|^{-1}\right)\widetilde{Y} \left(
  \frac{g_h}{g_t^{\Delta}},\left\{ g_u~ g_h^{ | y_u | / \Delta} \right\}
\right).  \ena The function $\wt{Y}(\cdot,\cdot)$ can be considered as a
constant in our analysis, whose exact value is $\wt{Y}(\cdot,\cdot)\equiv A_f= - \frac{4 \beta_c^2}{\pi}$.\\
The previous expression needs some clarification in order to extend it around
the magnetic axis.  Following the discussion of \cite{aarony}, we point out
that if we want $f(t,h_\ell)$ to be an analytic function of $t$ at fixed
$h_\ell$, a logarithmic term should appear in the expansion of
$Y(\cdot,\cdot)$.  In order to make the $\log | g_t |$ term disappear,
leaving us with the correct $\log | g_h |$, we have to extract a term like
\eqa x^2 |g_h|^\frac{2}{\Delta} \log|x|^{-1}A_f \ena from the expansion of
$Y(\cdot,\cdot)$ when $x \rightarrow 0$, where $x$ is the adimensional
variable $\frac{g_t}{ g_h^{1/\Delta}}$.  Hence the logarithmic contribution
has the following form~\footnote{In principle it is possible to include
  higher logarithmic powers in the free energy scaling function, However the
  leading order of the expansion of such terms are zero with a precision of
  $10^{-6}$ in our fits.}  \eqa f_{log}(t,h_\ell)=g_t^2 \log|g_h|^{-1}
\frac{A_f}{\Delta}.  \ena This gives the correct divergence of the specific
heat \cite{aarony} \eqa C\sim\frac{\partial^2 f}{\partial
  t^2}\sim2\frac{A_f}{\Delta} \log{|h_\ell|}+\dots.  \ena

With regard to the CFT approach one remark is in order:
the CFT derivation can generate only the singular term  ($f_{sing}$),while the bulk and the logarithmic contributions must be added separately.
 
The comparison of the two different approaches (CFT and Renormalization Group) 
enables us to understand the relation between the operators and the corrections they give rise to (for a careful discussion about this crucial point 
see \cite{Caselle:1999bx}). A relevant example is 
given by the stress-energy tensor: it is the only responsible for terms like $h^{32/15}$,
$t~h^{32/15}$, $t^2~h^{32/15}$.  
This will have fundamental implications in discussing the numerical data.
\section{Analysis of Results}
\label{sec:results}
The analysis of the data obtained by transfer matrix technique is done as follows:
\begin{itemize}
\item{To analyze the large amount of data, due to the large number of
    required terms in the expansion of the scaling functions, and in order to
    obtain sensible results in the fits, we fix as many terms as possible,
    resorting to both exact and high precision numerical results already
    known in the literature;}
\item{Fixing all known parameters in the scaling function and then fitting
    the data enables us to check their correctness in this particular regime
    ($\xi \rightarrow 0$, i.e. the neighborhood of the magnetic axis);}
\item{The high precision of our data enables us to conjecture the presence or
    the absence of contributions due to well identified sources, i.e. terms
    like $(A t + B t^2)~ h^{32/15}$, which are entirely due to the
    stress-energy tensor (the irrelevant fields $T {\bar T}$, $T^2$ and
    ${\bar T}$).}
\end{itemize}
This Section is devoted to developing such an analysis and to discuss our
results.
\subsection{Outline of the Numerical Computations}
We perform our numerical computations extracting the four larger
eigenvalues of the transfer matrix at given $t$, $h_\ell$, and $L$, the 
size of the transverse direction of the lattice. \\
The diagonalization method utilizes an iterative algorithm, due to \cite{RiNoRi},
that evaluates the highest-lying eigenvalues of the transfer matrix with arbitrary
high precision. The choice of an iterative algorithm lies in the reason that the typical size of the reduced transfer matrix is of order $10^5$ and the exact
diagonalization, even if partial, is an unobtainable computationally. \\
Our programs we run on 10 PCs equipped with Pentium III processors and
256 Mb of RAM for a total CPU time of about 2 months. \\
We performed our numerical computations for each available choice of the
three parameters $t$, $h_\ell$ and $L$ that we collected in table
\ref{par_val} for a total of 2600 different runs.

\begin{table}
\begin{center}
\begin{tabular}{l}
\hline
\hline
\multicolumn{1}{c}{$t$}\\
\hline
0.0130481\\0.0119726\\0.0108971\\0.0087367\\0.0076518\\0.0065668\\0.0043875\\0.0021985\\0.0010992\\0.0\\
\hline
\hline
\end{tabular}
\hspace{5mm}
\begin{tabular}{l}
\hline
\hline
\multicolumn{1}{c}{$h_\ell$}\\
\hline
0.01\\
0.02\\
0.03\\
0.04\\[1.5mm]
$\vdots$\\[1.5mm]
0.17\\
0.18\\
0.19\\
0.20\\
\hline
\hline
\end{tabular}
\hspace{5mm}
\begin{tabular}{l}
\hline
\hline
\multicolumn{1}{c}{$L$}\\
\hline
9\\
10\\
11\\
12\\[1.5mm]
$\vdots$\\[1.5mm]
18\\
19\\
20\\
21\\
\hline
\hline
\end{tabular}
\end{center}
\caption{Parameters value.}
\label{par_val}
\end{table}
\subsection{Infinite Volume Extrapolations}
After obtaining the values of the thermodynamical quantities at fixed $L$, we
would like to extrapolate them, taking the thermodynamic limit.  In order to
do this, we follow the method of \cite{Caselle:2000nn}, which we
outline briefly for sake of completeness. \\
The basic idea is that the behavior of any thermodynamic quantity in a
massive QFT will show an exponential decay as a function of $L$.  The task we
would like to achieve is to find the asymptotic value of the observables; in
order to reach this result, we must be able to subtract all the possible
exponential behaviors from our data.  To achieve we iteratively subtract the
exponential behavior from our data, removing both leading and subleading
corrections.  We define
\begin{equation}
\begin{split}
b^{i}(L_1-2) & = c~ \exp (-x(L_1 - 2))+ b^{i+1}(L_1)\\
b^{i}(L_1-1) & = c~ \exp (-x(L_1 - 1))+ b^{i+1}(L_1)\\
b^{i}(L_1) & = c~ \exp (-x(L_1))+ b^{i+1}(L_1)
\end{split}
\end{equation}
where $b^{0}(L_1)$ is the transfer matrix quantity. A step of
iteration is defined by solving the previous system with respect to 
$b^{i+1}(L_1)$, $c$ and $x$. The iteration chain stops when the
predicted value becomes numerically unstable. 
The taken result is the last stable prediction and its error is
evaluated from the variation with respect to previous step.  
\subsection{The Fitting Procedure}
In order to analyze our data we fit them with the scaling function 
previously obtained for a generic operator $\mathcal O$. \\
The fit is performed in three steps:
\begin{enumerate}
\item {
    We start fitting the data at fixed $h_\ell$ with a polynomial in the
    reduced temperature \eqa \mathcal{O} (t) q|_{h_\ell} - \mathcal{O} (0)
    |_{h_\ell} = \mathcal{A}^\mathcal{O}_1({h_\ell})~ t +
    \mathcal{A}^\mathcal{O}_2({h_\ell})~ t^2 +
    \mathcal{A}^\mathcal{O}_3({h_\ell})~t^3 +
    \mathcal{A}^\mathcal{O}_4({h_\ell})~ t^4 + \dots
\label{scaling_func_glob}
\ena where we found convenient to subtract the value of the observable at the
critical temperature; with this subtraction we are a position to discard
all the terms in the expansion that depend on $h_\ell$ only.  The results of
the fit are collected in a table displaying the values of
$\mathcal{A}_1^\mathcal{O}(h_\ell)$ and $\mathcal{A}_2^\mathcal{O}(h_\ell)$ as a functions of $h_\ell$.}
\item{
    By means of the expansion for the scaling function, we can fit the
    functions $\mathcal{A}_1^\mathcal{O}(h_\ell)$ and $\mathcal{A}_2^\mathcal{O}(h_\ell)$ against the magnetic field
    $h_\ell$. In this way we were able to obtain the amplitudes of the
    corrections to the observable $\mathcal{O}$ at first and second order in
    the reduced temperature $t$. }
\item{
    We use all predicted values of the parameters of the fit to calculate the
    $\chi^2$ with a unique fitting function for all the data (the
    coefficients $\mathcal{A}_3^\mathcal{O}(h_\ell)$ and $\mathcal{A}_4^\mathcal{O}(h_\ell)$ of step 1 are only useful to the determination
    of the $\chi^2$ of the entire set of data).}
\end{enumerate}
The list of requirements a fit must fulfill to be accepted is the following:
\begin{itemize}
\item{The reduced $\chi^2$ of the fit had to be of order unity, i.e. we required it to have 
a con\-fi\-dence level larger than $30 \% $.}
\item{In order to be included in the fitting function, the subleading terms must have an 
amplitude larger than the corresponding error.}
\item{The number of degrees of freedom must be larger than 5.}
\end{itemize}
The previous requirements are very hard to fulfill at the same time; this way
we also take into account the systematic errors due to extrapolations.
\subsection{Numerical Results}
\subsubsection{Determination of $e_h$}
The least precise of the constants collected in the appendix \ref{numbers} is
$e_h$.  The high precision of our data enables us to make an attempt to
improve this estimate.  Following the same procedure of
\cite{Caselle:1999bx}, and fixing all the known terms in the scaling function
we are able to give the following result \eqa e_h = -0.007298(3) \ena where
the data we used as input are collected in appendix \ref{numbers}.
\subsubsection{Free Energy}
The analysis of the free energy can be performed in four steps. 
\begin{enumerate}

\item[Firstly,]{we fit the the term proportional to $t$ with the following function
\begin{equation}
\begin{split}
&-0.623226~ -0.511645~h_\ell^\frac{8}{15}+
0.329993~h_\ell^\frac{16}{15}
 -0.0427280~h_\ell^2+\\
&-0.0107535~h_\ell^2 \log |h_\ell|+
h_\ell^\frac{32}{15}~\mathcal{K}_0+
0.0184589~h_\ell^\frac{38}{15}+h_\ell^\frac{40}{15}~\mathcal{K}_1+\\
&+
0.0001605~h_\ell^\frac{46}{15}+
h_\ell^\frac{48}{15}~\mathcal{K}_2+
h_\ell^\frac{52}{15}~\mathcal{K}_3+\dots.
\end{split}
\end{equation}
As we explain in the appendix \ref{numbers} we set $Y^{(2,1)}(0,0)=0$ according to the results of \cite{Gart} and \cite{gutty}, moreover
we can found $Y^{(1,1)}(0,0)$ compatible with zero within the error in all our fits.\\
The fit shows a non-zero correction due to the stress-energy tensor, in fact we are able to estimate the value of $\mathcal{K}_0$
\eqa
-0.05828 <\mathcal{K}_0<-0.05820.
\ena
where the expression of $\mathcal{K}_0$ in terms of the scaling fields is given by
\eqa
\mathcal{K}_0 = \left(a_u + 0.664773 u \right) Y^{(0,1)}(0,0).
\ena
This is quite an interesting result, in fact it is known that in both the 
integrable perturbation of the critical Ising model, the amplitude of this term is compatible with zero.}
\item[Secondly,]{we perform the fit of $t^2$ term with the following function
\begin{equation}
\begin{split}
&\mathcal{K}_4-0.131877~\log |h_\ell| -0.244467~h_\ell^\frac{8}{15} + 0.299064~h_\ell^\frac{16}{15} +\\&+~ \mathcal{K}_5~h_\ell^\frac{22}{15} -0.0318458 ~h_\ell^2 + 0.0100527~h_\ell^2~\log |h_\ell| + h_\ell^\frac{32}{15}~\mathcal{K}_6+\dots.
\end{split}
\end{equation}
As previously stated we put $Y^{(2,1)}(0,0)=Y^{(1,1)}(0,0)=0$, and we found that leading correction is 
\eqa
0.03655 <\mathcal{K}_4<0.03657.
\ena
The expression of $\mathcal{K}_4$ in terms of the scaling fields is:
\eqa
\mathcal{K}_4 = 0.481290866 + f^b_{2,0}
\ena
in this way we are able to give an estimate for the bulk term
\eqa
-0.44474 <f^b_{2,0}<-0.44472.
\label{fbulk20}
\ena
Furthermore, we can clearly see that the 
term due to the stress-energy tensor $\mathcal{K}_6$ (which comes with the power $h_\ell^{32/15}$) is different from zero, as for the previous case.\\
As discussed in \cite{Caselle:1999bx,Caselle:2002cs} it is difficult to have reliable estimates for the amplitudes of the subleading terms\footnote{As discussed in \cite{Caselle:1999bx}, notwithstanding systematic errors, one can give also a rough estimate of the subleading amplitudes. Following the same route
we are able to find
\eqa
\label{k5footnote}
-0.721 < \mathcal{K}_5 < -0.717.
\ena} due to large systematic deviations induced by the uncertainties in the leading corrections, nevertheless we can still assert that they are different from zero.}
\item[Thirdly,]{we check that within the precision of our computation, the function 
$\wt{Y}(\cdot,\cdot)$ is a constant. In fact, if we expand it in Taylor series, we find 
that the first contribution to the scaling function comes in the $t^2$ term and has the
following form
\eqa
\wt{Y}^{(1,0)}(0,0)~ \log | h_\ell | ~ h_\ell^{22/15}.
\ena  
}
If we perform the fit with this new contribution, we find that its amplitude is compatible 
with zero. Hence, for our purposes it is safe to consider $\wt{Y}(\cdot,\cdot)$ as a constant. 
\item[Fourthly,]{we compute the $\chi^2$ with the global scaling function (Eq.~\ref{scaling_func_glob}) both on $t$ and $h_\ell$ utilizing the value of the constants predicted in the previous steps, in order to verify the correctness of the fitting procedure. 
}
\end{enumerate}
It is important to mention  that in the fitting functions we use the
 known values of the constants of Appendix~\ref{numbers}. It is a non-trivial test on the validity of both our results, and the well known values we used.
\subsubsection{Magnetization}
\begin{enumerate}
\item[Firstly,]{we write down the correction proportional to the $t$ term
\begin{equation}
\begin{split}
&-\frac{0.2728775}{h_\ell^\frac{7}{15}}
+   0.3519930~h_\ell^\frac{1}{15} 
-   0.0747026~h_\ell
+   0.0215069~h_\ell \log |h_\ell|+\\
&+  h_\ell^\frac{17}{15} \mathcal{H}_0
+   0.0467625~h_\ell^\frac{23}{15}
+   h_\ell^\frac{5}{3} \mathcal{H}_1
-   0.0004923~h_\ell^\frac{31}{15}
+   h_\ell^\frac{11}{5} \mathcal{H}_2+\\
&+   h_\ell^\frac{37}{15} \mathcal{H}_3
+   h_\ell^\frac{41}{15} \mathcal{H}_4+\dots
\end{split}
\end{equation}
where all the known quantities are taken into account. In particular we also
impose the constraints we found in the analysis of the free energy,
i.e. $Y^{(2,1)}(0,0)=Y^{(1,1)}(0,0)=0$, and we find perfect agreement also in this case. \\
The leading correction is due to the stress-energy tensor, as we expected from the analysis of the free energy, and its amplitude is given by
\eqa
-0.124 < \mathcal{H}_0 < -0.122
\ena
which is compatible with $\mathcal{H}_0 \equiv\frac{32}{15}~\mathcal{K}_0$ within our errors.
}
\item[Secondly,]{the contribution due to $t^2$ is given by
\begin{equation}
\begin{split}
&-\frac{0.1318769}{h_\ell}
-\frac{0.1303825}{h_\ell^\frac{7}{15}} + 
 0.3190016~h_\ell^\frac{1}{15}
+\frac{22}{15}~ h_\ell^\frac{7}{15}~\mathcal{H}_5
-0.0536389~h_\ell+\\
&+0.0201054~h_\ell~\log| h_\ell|
+h_\ell^\frac{17}{15}~\mathcal{H}_6+\dots
\end{split}
\end{equation}
and is obtained following the same strategy as before. This order also shows a non-zero contribution due to stress-energy tensor, i.e. $\mathcal{H}_6$. \\
The amplitude of leading order correction is given by
\eqa
 -0.723 < \mathcal{H}_5 < -0.718.
\ena 
We can check that it is in reasonable agreement with the subleading amplitude
of the free energy $\mathcal{H}_5 \equiv \mathcal{K}_5$ (see (\ref{k5footnote})).}
\end{enumerate}
\subsubsection{Internal Energy}
\begin{enumerate}
\item[Firstly,]{the scaling function for the $t$ term is
\begin{equation}
\begin{split}
&\mathcal{W}_0
+0.2992532 \log |h_\ell|
+0.5547414~h_\ell^\frac{8}{15}
+h_\ell^\frac{16}{15} \mathcal{W}_1
+h_\ell^\frac{22}{15} \mathcal{W}_2
-0.0918396~h_\ell^2+\\
&+h_\ell^\frac{32}{15} \mathcal{W}_3+h_\ell^\frac{38}{15} \mathcal{W}_4
-0.0228115~h_\ell^2~\log |h_\ell|+\dots
\end{split}
\end{equation}
the predicted value for the leading contribution is
\eqa
-0.1659<\mathcal{W}_0<-0.1657
\ena
which expressed in term of scaling function is $\mathcal{W}_0\equiv (-2.1842763+ 4.5383706~f^b_{2,0})$, the value of $f^b_{2,0}$ is consistent with the previous estimate (\ref{fbulk20}).
As before the fit is compatible with the constraint $Y^{(2,1)}(0,0)=Y^{(1,1)}(0,0)=0$.}
\item[Secondly,]{the scaling function for the $t^2$ term is
\begin{equation}
\begin{split}
& \frac{\mathcal{W}_5}{h_\ell^\frac{8}{15}} +
\mathcal{W}_6
-0.2797532~\log |h_\ell|+
h_\ell^\frac{8}{15}~\mathcal{W}_7 + 
h_\ell^\frac{14}{15}~\mathcal{W}_6 +
h_\ell^\frac{16}{15}~\mathcal{W}_8 +\\
&- 0.0727482~h_\ell^\frac{22}{15} \mathcal{W}_5+\dots
\end{split}
\end{equation}
the predicted value for the leading contribution is
\eqa
-0.414<\mathcal{W}_5<-0.408
\ena
where $\mathcal{W}_5\equiv 0.5672963~Y^{(3,0)}(0,0)$,
and again the fit is compatible with the constraint $Y^{(2,1)}(0,0)=Y^{(1,1)}(0,0)=0$.
}
\end{enumerate}
\subsubsection{Susceptibility}
The form of the scaling function is in total agreement with the computed
values of the this observable. However no new predictions are available.
\subsubsection{Mass spectrum}
In order to check the results of the FFPT proposed by \cite{Delfino:1996xp}, we performed the fit on the square of the first three masses of the theory, and computed the following ratio 
\eqa
\frac{\delta m_i^2}{\delta m_1^2}=\frac{m(t)_i^2-m(0)_i^2}{m(t)_1^2-m(0)_1^2}
\ena
where $m(t)_i^2$ are the perturbed masses, and $m(0)_i^2$ the unperturbed ones.\\
The results of our analysis are reported below:
\begin{equation}
\begin{split}
\frac{\delta m_2^2}{\delta m_1^2}&=1.393(17)\\
\frac{\delta m_3^2}{\delta m_1^2}&=3.16(30).
\end{split}
\end{equation}
These results are in perfect agreement with the theoretical prediction.  In fact, since the deviation from the integrable model is small, it is correct to write $\delta m_i^2 = 2 m(0)_i \delta m_i$, and finally we have
\begin{equation}
\begin{split}
\frac{\delta m_2}{\delta m_1}&=0.86(1)\\
\frac{\delta m_3}{\delta m_1}&=1.58(15).
\end{split}
\end{equation}

\subsubsection{Magnetization Overlap}
The magnetic overlap $| F_1^\sigma|^2$ can also be analyzed as
before. We are able to find the leading order corrections of terms proportional to $t$
\begin{equation}
\begin{split}
\frac{\mathcal{R}_0}{h^\frac{8}{15}}
+ h^\frac{2}{15} \mathcal{R}_1
+ h^\frac{8}{15} \mathcal{R}_2
+ h^\frac{14}{15} \mathcal{R}_3
+ h^\frac{16}{15} \mathcal{R}_4
+ h^\frac{22}{15} \mathcal{R}_5
+ h^\frac{8}{5} \mathcal{R}_6
+ h^\frac{28}{15} \mathcal{R}_7
+ h^2 \mathcal{R}_8+\dots
\end{split}
\end{equation}
and $t^2$
\begin{equation}
\begin{split}
\frac{\mathcal{R}_9}{h^\frac{16}{15}}
+ \frac{\mathcal{R}_{10}}{h^\frac{8}{15}}
+ \frac{\mathcal{R}_{11}}{h^\frac{2}{15}}
+ h^\frac{2}{5} \mathcal{R}_{12}
+ h^\frac{8}{15} \mathcal{R}_{13}
+ h^\frac{14}{15} \mathcal{R}_{14}+\dots.
\end{split}
\end{equation}
The numerical values are
\begin{equation}
\begin{split}
0.628<\mathcal{R}_{0}<0.631\\
0.661<\mathcal{R}_{9}<0.664
\end{split}
\end{equation}
It would be interesting to calculate the same corrections on theoretical
grounds (at least for the $t$ correction) in order to make a comparison.
\section{Discussion and Conclusions}
\label{sec:discandconcl}
In this paper we studied the effect of a mixed relevant perturbation on the
Ising model using the Transfer Matrix technique. We consider the neighborhood
of the magnetic axis in the limit $\xi \ll 1 $, where $\xi $ is the
adimensional parameter $t/h^{8/15}$.

We have concentrated our efforts on the following areas
\begin{itemize}
\item{We calculated the scaling functions (appendix \ref{scalingfunc})
    applying a CFT approach to the problem.}
\item{We used all known predictions about the behavior of scaling functions
    (see appendix \ref{numbers}), and verified that they all agree with our
    data.}
\item{Being able to identify the contribution of secondary fields to the
    scaling function, we predicted that there is a {\it non zero}
    contribution due to the stress-energy tensor, and we evaluate it. In our
    opinion this a quite interesting result, because it is known that the
    contribution of these particular secondary fields is zero if we study the
    model with only one relevant perturbation.}
\item{We obtained estimates of several amplitudes never predicted by any
    other analytical method before.}
\item{We calculated the correction to the Zamolodchikov mass spectrum of
    the Ising model with a magnetic field, and we found perfect agreement
    with Delfino et al. \cite{Delfino:1996xp}.}
\end{itemize}
There are two possible developments of this work: We can use our data to
improve the knowledge of the equation of state of the Ising model, in order
to map all the possible regime of perturbations.  We can extend the analysis
of \cite{Caselle:2002cs} to study the effect of mixed perturbation on the
finite temperature results.
\section*{Acknowledgments}
We thank M.Caselle and H.Hasenbusch for useful suggestions, and several useful discussions. 
\clearpage
\appendix
\section{Known Numbers}
\label{numbers}
It is known that it is possible to write the free energy of the model in terms of nonlinear scaling fields \cite{Wegner_1976}.

The scaling fields are analytic functions of $t$ and $h_\ell$ respecting the $\mathbb{Z}_2$ parity of $h_\ell$. Their Taylor expansions are expected to be
\begin{equation}
\begin{split}
g_t(t,h_\ell) &= t + b_t h_\ell^2 + c_t t^2 + d_t t^3 + e_t t h_\ell^2 +
f_t t^4 + g_t h_\ell^4 + h_t t^2 h_\ell^2 + \dots\\
g_h(t,h_\ell) &=  h_\ell (1+ c_h t + d_h t^2 + e_h h_\ell^2 + f_h t^3
+ g_h t h_\ell^2) + \\
g_u(t,h_\ell) &= u + a_u t +b_u h_\ell^2 + c_u t^2 + d_u t^3 + e_u t
h_\ell^2 + f_u h_\ell^4  + l_u t^2 h_\ell^2 + \dots
\end{split}
\end{equation}
Here we report all the analytically known coefficients \cite{gutty,Caselle:2000bj,Nickel,Salas:1999qh}
\begin{gather}
c_h=\frac{\beta_c}{\sqrt{2}},  \hskip1cm
d_h=\frac{23 \beta_c^2}{16},   \hskip1cm
f_h=\frac{191 \beta_c^3}{48\sqrt{2}},\nonumber \\
c_t=\frac{\beta_c}{\sqrt{2}},\hskip1cm
d_t=\frac{7 \beta_c^2}{6}, \hskip1cm
f_t=\frac{17 \beta_c^3}{6\sqrt{2}},\\
e_t = b_t \beta_c \sqrt{2}, \qquad \qquad b_t = - \frac{E_0 \pi}{16 \beta_c^2}\nonumber
\end{gather}
where
\eqa
E_0=0.0403255003\dots
\label{E0}
\ena
From the analysis of the model at the critical temperature and $h_\ell \neq 0$,
 we obtain a new estimate for $e_h$
\eqa
e_h = -0.007298(3).
\ena
For the free energy, we use  also the coefficient reported in the following
\begin{equation}
\begin{split}
Y(0,0)&=0.99279949\dots\\
Y^{(1,0)}(0,0)&=A^l_{f,2}/b_t=-0.511645336\dots\\
A^l_{f,2}&=\frac{A^l_{E} \pi}{Y(0,0) 8 \beta_c}=0.0208602\dots\\
f_b&=\frac{2}{\pi} G+\frac{1}{2}\log 2\\
A^l_{f,b}&=-0.0524442\dots
\end{split}
\end{equation}
where $G$ is the Catalan constant and the $A$'s are defined in \cite{Caselle:1999bx}.\\
A high precision study of the thermal perturbation have been performed by Orrick et al. in  \cite{gutty}, from this work we are able to extract a set of parameters for our scaling functions.\\
In particular we are able to observe that the known coefficients for both thermal and magnetic perturbations are exactly consistent, and we can extend the knowledge about the thermal coefficients to gain predictions about the unknown magnetic terms.\\
To achieve these results we compare the scaling function of the susceptibility along the thermal axis with the high precision estimates of \cite{gutty} and \cite{Gart}, in this way we obtained
\begin{equation}
\begin{split}
Y^{(2,0)}(0,0) = 0.9625817322 \dots\\
f^b_{0,2} \equiv A^l_{f,b}~ Y(0,0) = -0.05206662255 \dots\\
f^b_{1,2} = -0.00348278\dots\\
f^b_{2,2} = 0.000528775 \dots.
\end{split}
\end{equation}
These results are obtained for the thermal perturbation in the regime $t\neq0,h_\ell=0$, we observe that these results are valid also in our regime of interest, because the analytic continuation of Section~\ref{4.2.1} do not affect similar kinds of terms, in total agreement with our fits.\\
Furthermore we obtain also the relations:
\begin{equation}
\begin{split}
Y^{(2,1)}~(0,0) (a_u+ 0.701128 u) =0\\
Y^{(2,1)}~(0,0) u =0
\end{split}
\end{equation}
In order to fulfill the above requirements, we set $Y^{(2,1)}~(0,0)=0$ because
the other choice $u =0$, $a_u =0$ was not consistent with our fits.

Finally, from Onsager's exact solution \cite{Onsager:1943jn} we shall extract
\eqa
f^b_{1,0} = -0.623226\dots . 
\ena                     
\section{Scaling Functions}
\label{scalingfunc}
Here we report some of the scaling functions we obtained.
We remark that the functions have been evaluated to higher order of the expansion reported here ($O(h_\ell^5) O(t^5)$).\\
Free Energy:
\eqa
f(t) |_{h_\ell} - f(0) |_{h_\ell} = \mathcal{A}^f_1({h_\ell})~ t +  \mathcal{A}^f_2({h_\ell})~ t^2 +  \dots
\ena
\begin{equation}
\begin{split}
\mathcal{A}^f_1(h_\ell) & = 
{f^b_{1,0}} + Y^{(1,0)}(0,0)\,h_\ell^{\frac{8}{15}} + \frac{16}{15}\,{c_h}\,Y(0,0)\,h_\ell^{\frac{16}{15}} + 
  u\,Y^{(1,1)}(0,0)\,h_\ell^{\frac{8}{5}} +  \\
& +  \left(- \frac{16}{15}\,A_f\,\log | h_\ell |\,{b_t} + {f^b_{1,2}} + 
     {b_t}\,Y^{(2,1)}(0,0) \right) \,h_\ell^2 + \dots
\end{split}
\end{equation}
\begin{equation}
\begin{split}
\mathcal{A}^f_2(h_\ell) & = 
\left(-\frac{8}{15}\,A_f\,\log | h_\ell | + {f^b_{0,2}} + \frac{1}{2} Y^{(2,0)}(0,0) \right) + 
  \left( \frac{8}{15}\,{c_h}\,Y^{(1,0)}(0,0)  \, \right.+ \\
&\left.+ \,
{c_t}\,Y^{(1,0)}(0,0)\hspace{-4mm} \phantom{\frac{1}{1}}  \right) \,h_\ell^{\frac{8}{15}} + 
  \left( \frac{8}{225}\,{{c_h}}^2\,Y(0,0) + \frac{16}{15}\,{d_h}\,Y(0,0) + \frac{1}{2}u\,Y^{(2,1)}(0,0) \right) \, h_\ell^{\frac{16}{15}} +  \\
& + 
\frac{1}{2}{b_t}\,Y^{(3,0)}(0,0)\,h_\ell^{\frac{22}{15}} + 
  \left( {a_u}\,Y^{(1,1)}(0,0) + \frac{8}{5}\,u\,{c_h}\,Y^{(1,1)}(0,0) + u\,{c_t}\,Y^{(1,1)}(0,0) \right) \, h_\ell^{\frac{8}{5}} +  \\
& + 
\left(- \frac{16}{15}\,A_f\,{b_t}\,{c_h} - 
     \frac{16}{15}\,A_f\,\log | h_\ell |\,{b_t}\,{c_t} - \frac{8}{15}\,A_f\,{e_h} - 
     \frac{16}{15}\,A_f\,\log | h_\ell |\,{e}_t + {f^b_{2,2}} \right. + \\
&\left. + \, {b_t}\,{c_t}\,Y^{(2,0)}(0,0) + 
     {e_t}\,Y^{(2,0)}(0,0) \hspace{-4mm} \phantom{\frac{1}{1}} \right) \,h_\ell^2 
+ \dots
\end{split}
\end{equation}
Internal Energy:
\eqa
E(t) |_{h_\ell} - E(0) |_{h_\ell} = \mathcal{A}^E_1({h_\ell})~ t +  \mathcal{A}^E_2({h_\ell})~ t^2 +  \dots
\ena
\begin{equation}
\begin{split}
\mathcal{A}^E_1(h_\ell) & = \left(-\frac{16}{15}\,A_f\,\log | h_\ell |\,T_c + 2\,{f^b_{2,0}}\,{T_c} + {T_c}\,Y^{(2,0)}(0,0)\right) +\\
&+  \left( \frac{16}{15}\,{c_h}\,{T_c}\,Y^{(1,0)}(0,0) + 2\,{c_t}\,{T_c}\,Y^{(1,0)}(0,0) \right) \,h_\ell^{\frac{8}{15}} + \\
&+   \left( \frac{16}{225}\,{{c_h}}^2\,{T_c}\,Y(0,0) + \frac{32}{15}\,{d_h}\,{T_c}\,Y(0,0) + u\,{T_c}\,Y^{(2,1)}(0,0) \right) \,
   h_\ell^{\frac{16}{15}} + \\
&+\,  {b_t}\,{T_c}\,Y^{(3,0)}(0,0)\,h_\ell^{\frac{22}{15}} + 
  \left( 2\,{a_u}\,{T_c}\,Y^{(1,1)}(0,0) \phantom{\frac{1}{1}}\right.+\\
&\left.+\, \frac{16}{5}\,u\,{c_h}\,{T_c}\,Y^{(1,1)}(0,0) + 
     2\,u\,{c_t}\,{T_c}\,Y^{(1,1)}(0,0) \right) \,h_\ell^{\frac{8}{5}} + \dots
\end{split}
\end{equation}
\begin{equation}
\begin{split}
\mathcal{A}^E_2(h_\ell) &= \frac{{T_c}}{2\,h_\ell^{\frac{8}{15}} }\,Y^{(3,0)}(0,0) + 
  \left(-\frac{8}{5}\,A_f\,{c_h}\,{T_c} - \frac{16}{5}\,A_f\,\log | h_\ell |\,c_t \,{T_c} + 3\,{f^b_{3,0}}\,{T_c} +\right.\\
&+\, 3\,{c_t}\,{T_c}\,Y^{(2,0)}(0,0)\left. \hspace{-4.5mm}\phantom{\frac{1}{1}} \right) +\left( -\frac{28}{75}\,{{c_h}}^2\,{T_c}\,Y^{(1,0)}(0,0) + 
     \frac{8}{5}\,{c_h}\,{c_t}\,{T_c}\,Y^{(1,0)}(0,0) +\right.\\
&+\,\left. \frac{8}{5}\,{d_h}\,{T_c}\,Y^{(1,0)}(0,0) +3\,{d_t}\,{T_c}\,Y^{(1,0)}(0,0) + \frac{1}{2}u\,{T_c}\,Y^{(3,1)}(0,0) \right) 
\,h_\ell^{\frac{8}{15}} +\\
&\,+\frac{1}{2}{b_t}\,{T_c}\,Y^{(4,0)}(0,0)\,h_\ell^{\frac{14}{15}} +\left( -\frac{112}{3375}\,{{c_h}}^3\,{T_c}\,Y(0,0) + \frac{16}{75}\,{c_h}\,{d_h}\,{T_c}\,Y(0,0) +\right.\\
&+\,\left. \frac{16}{5}\,{f_h}\,{T_c}\,Y(0,0) + \frac{3}{2}\,{a_u}\,{T_c}\,Y^{(2,1)}(0,0) +\frac{8}{5}\,u\,{c_h}\,{T_c}\,Y^{(2,1)}(0,0) +\right.\\
&+\,\left. 3\,u\,{c_t}\,{T_c}\,Y^{(2,1)}(0,0) \hspace{-4mm}\phantom{\frac{1}{1}}\right) \,h_\ell^{\frac{16}{15}} + \left( -\frac{4}{5}\,{b_t}\,{c_h}\,{T_c}\,Y^{(3,0)}(0,0) + 3\,{b_t}\,{c_t}\,{T_c}\,Y^{(3,0)}(0,0) +\right.\\
&-\,\left. \frac{4}{15}\,{e_h}\,{T_c}\,Y^{(3,0)}(0,0) + \frac{3}{2}\,{e_t}\,{T_c}\,Y^{(3,0)}(0,0) \right) \,h_\ell^{\frac{22}{15}} +  \left( \frac{24}{5}\,{a_u}\,{c_h}\,{T_c}\,Y^{(1,1)}(0,0) +\right.\\
&+\,\left. \frac{36}{25}\,u\,{{c_h}}^2\,{T_c}\,Y^{(1,1)}(0,0) + 
     3\,{a_u}\,{c_t}\,{T_c}\,Y^{(1,1)}(0,0) +\frac{24}{5}\,u\,{c_h}\,{c_t}\,{T_c}\,Y^{(1,1)}(0,0) + \right.\\
&+\,\left. 3\,{c_u}\,{T_c}\,Y^{(1,1)}(0,0) +\frac{24}{5}\,u\,{d_h}\,{T_c}\,Y^{(1,1)}(0,0) + 3\,u\,{d_t}\,{T_c}\,Y^{(1,1)}(0,0) +\right.\\
&+\,\left. \frac{1}{4}u^2\,{T_c}\,Y^{(3,2)}(0,0) \right) \,h_\ell^{\frac{8}{5}} + \dots
\end{split}
\end{equation}
\section{Example of numerical data}
Here we report, for sake of completeness, an example of the data we used to fit.\\
We refer to value of $\beta=0.4373$. 
\begin{table}
\begin{center}
\begin{tabular}{ll}
\hline
\hline
&\\[-5mm]
$h_\ell$&$E(t,h_\ell)$\\
\hline
\hline
&\\[-5mm]
$0.01$&$0.035901(3)$\\
$0.02$&$0.0569877(2)$\\
$0.03$&$0.07237192(2)$\\
$0.04$&$0.084838424(2)$\\
$0.05$&$0.0954482541(1)$\\
$0.06$&$0.1047420407(1)$\\
$0.07$&$0.11303997732(1)$\\
$0.08$&$0.12055040575(1)$\\
$0.09$&$0.12741797596(1)$\\
$0.10$&$0.13374801551(1)$\\
$0.11$&$0.13962007086(2)$\\
$0.12$&$0.145095979793(2)$\\
$0.13$&$0.150224955365(1)$\\
$0.14$&$0.155046932292(1)$\\
$0.15$&$0.159594851224(1)$\\
$0.16$&$0.16389626622(1)$\\
$0.17$&$0.167974505628(1)$\\
$0.18$&$0.171849529388(1)$\\
$0.19$&$0.17553857454(1)$\\
$0.20$&$0.179056649653(1)$\\
\hline
\hline
\end{tabular}
\hspace{-3mm}
\begin{tabular}{l}
\hline
\hline
\\[-5mm]
$f(t,h_\ell)$\\
\hline
\hline
\\[-5mm]
$0.931944971(5)$\\
$0.939788377(1)$\\
$0.94794151643(2)$\\
$0.95629353642(1)$\\
$0.96479217393(1)$\\
$0.97340669611(1)$\\
$0.98211679357(1)$\\
$0.99090802119(1)$\\
$0.999769568339(1)$\\
$1.008693035843(1)$\\
$1.017671707919(1)$\\
$1.026700091153(1)$\\
$1.035773608172(1)$\\
$1.044888386026(1)$\\
$1.054041105248(1)$\\
$1.06322888925(1)$\\
$1.072449221416(1)$\\
$1.081699881718(1)$\\
$1.090978897417(1)$\\
$1.100284504134(1)$\\
\hline
\hline
\end{tabular}
\hspace{-3mm}
\begin{tabular}{l}
\hline
\hline
\\[-5mm]
$M(t,h_\ell)$\\
\hline
\hline
\\[-5mm]
$0.76120(4)$\\
$0.802572(1)$\\
$0.82640939(1)$\\
$0.843168904(1)$\\
$0.8560620814(1)$\\
$0.8665107103(1)$\\
$0.8752714873(1)$\\
$0.8827957916(1)$\\
$0.88937481546(1)$\\
$0.89520749914(1)$\\
$0.90043585646(1)$\\
$0.905164845833(3)$\\
$0.909474255678(2)$\\
$0.913426172207(2)$\\
$0.917069865355(1)$\\
$0.920445095734(1)$\\
$0.923584418311(1)$\\
$0.926514827477(1)$\\
$0.92925895744(1)$\\
$0.931835974906(1)$\\
\hline
\hline
\end{tabular}
\hspace{-3mm}
\begin{tabular}{l}
\hline
\hline
\\[-5mm]
$\chi(t,h_\ell)$\\
\hline
\hline
\\[-5mm]
\\
\\
\\
$1.45013(2)$\\
$1.150734(1)$\\
$0.9510868(1)$\\
$0.808372(1)$\\
$0.701236(1)$\\
$0.6178284(1)$\\
$0.5510433(1)$\\
$0.4963581(1)$\\
$0.4507573(1)$\\
$0.4121515(1)$\\
$0.3790476(1)$\\
$0.35034994(1)$\\
$0.32523603(1)$\\
$0.303076426(1)$\\
$0.283381176(1)$\\
$0.2657631501(1)$\\
$0.2499123945(1)$\\
\hline
\hline
\end{tabular}
\end{center}
\end{table}

\begin{table}
\begin{center}
\begin{tabular}{ll}
\hline
\hline
&\\[-5mm]
$h_\ell$&$1/m_1(t,h_\ell)$\\
\hline
\hline
&\\[-5mm]
$0.03$&$1.66609(4)$\\
$0.04$&$1.42446(2)$\\
$0.05$&$1.262217(1)$\\
$0.06$&$1.1438966(2)$\\
$0.07$&$1.0528411(3)$\\
$0.08$&$0.980049866(1)$\\
$0.09$&$0.920186381(1)$\\
$0.10$&$0.8698623782(3)$\\
$0.11$&$0.82681092694(1)$\\
$0.12$&$0.789451203443(1)$\\
$0.13$&$0.756643166957(1)$\\
$0.14$&$0.727541458508(3)$\\
$0.15$&$0.701504369616(1)$\\
$0.16$&$0.678034937921(1)$\\
$0.17$&$0.656741586699(1)$\\
$0.18$&$0.637311086114(2)$\\
$0.19$&$0.619489526443(1)$\\
$0.20$&$0.603068643116(1)$\\
\hline
\hline
\end{tabular}
\hspace{-3mm}
\begin{tabular}{l}
\hline
\hline
\\[-5mm]
$1/m_2(t,h_\ell)$\\
\hline
\hline
\\[-5mm]
\\
$0.86(2)$\\
$0.775(2)$\\
$0.704762(5)$\\
$0.64979(3)$\\
$0.60587(2)$\\
$0.569787(5)$\\
$0.539467(1)$\\
$0.5135465(5)$\\
$0.4910712(1)$\\
$0.47135216(1)$\\
$0.453877992(1)$\\
$0.4382602553(3)$\\
$0.424197853(2)$\\
$0.41145329066(3)$\\
$0.39983637405(1)$\\
$0.389192731029(1)$\\
$0.379395558407(1)$\\
\hline
\hline
\end{tabular}
\hspace{-3mm}
\begin{tabular}{l}
\hline
\hline
\\[-5mm]
$1/m_3(t,h_\ell)$\\
\hline
\hline
\\
\\[-5mm]
\\
\\
\\
$0.5334(3)$\\
$0.4973(3)$\\
$0.4676(3)$\\
$0.4417(3)$\\
$0.4203(2)$\\
$0.4019(1)$\\
$0.38589(7)$\\
$0.37178(2)$\\
$0.35921(1)$\\
$0.347936(2)$\\
$0.3377387(2)$\\
$0.3284631(3)$\\
$0.31998(1)$\\
$0.312184(1)$\\
\hline
\hline
\end{tabular}
\hspace{-3mm}
\begin{tabular}{l}
\hline
\hline
\\[-5mm]
$|F_1^{\sigma}(t,h_\ell)|^2$\\
\hline
\hline
\\
\\[-5mm]
$0.413(1)$\\
$0.403172(4)$\\
$0.39427(1)$\\
$0.3859807(2)$\\
$0.37813145(1)$\\
$0.3706210923(3)$\\
$0.3633856671(3)$\\
$0.356382554(1)$\\
$0.3495820201(7)$\\
$0.34296250455(5)$\\
$0.33650783473(4)$\\
$0.3302055233(2)$\\
$0.324045673287(8)$\\
$0.318020257612(3)$\\
$0.31212262996(1)$\\
$0.306347184601(2)$\\
$0.300689114577(1)$\\
\hline
\hline
\end{tabular}
\end{center}
\end{table}

\end{document}